\documentclass[superscriptaddress,prl,twocolumn,tightenlines,10pt,a4paper]{revtex4}%

\usepackage{bbm}
\usepackage{amsfonts}
\usepackage{amssymb}
\usepackage{graphicx}
\usepackage{subfigure}
\usepackage{savesym}
\usepackage{amsmath}
\usepackage{txfonts}
\usepackage{multirow}
\usepackage{epstopdf}
\usepackage{color}
\usepackage{umoline}

\newcommand{\ket}[1]{\vert#1\rangle}
\newcommand{\bra}[1]{\langle#1\vert}

\usepackage{pbox}

\newcommand{\spinup}{|\!\uparrow\rangle}
\newcommand{\spindown}{|\!\downarrow\rangle}
\newcommand{\ue}{\mathrm{e}}

\begin{document}

\title{Observation of entanglement propagation in a quantum many-body system}

\author{P. Jurcevic}
\thanks{These authors contributed equally to this work.}
\author{B. P. Lanyon}
	\thanks{These authors contributed equally to this work.}

	\affiliation{Institut f\"ur Quantenoptik und Quanteninformation,\\
	\"Osterreichische Akademie der Wissenschaften, Technikerstr. 21A, 6020 Innsbruck,
	Austria}
	\affiliation{
	Institut f\"ur Experimentalphysik, Universit\"at Innsbruck,
	Technikerstr. 25, 6020 Innsbruck, Austria}

\author{P. Hauke}

	\affiliation{Institut f\"ur Quantenoptik und Quanteninformation,\\
	\"Osterreichische Akademie der Wissenschaften, Technikerstr. 21A, 6020 Innsbruck,
	Austria}
	\affiliation{
	Institut f\"ur Theoretische Physik, Universit\"at Innsbruck,
	Technikerstr. 25, 6020 Innsbruck, Austria}

\author{C. Hempel}
 
	\affiliation{Institut f\"ur Quantenoptik und Quanteninformation,\\
	\"Osterreichische Akademie der Wissenschaften, Technikerstr. 21A, 6020 Innsbruck,
	Austria}
	\affiliation{
	Institut f\"ur Experimentalphysik, Universit\"at Innsbruck,
	Technikerstr. 25, 6020 Innsbruck, Austria}

\author{P. Zoller}

	\affiliation{Institut f\"ur Quantenoptik und Quanteninformation,\\
	\"Osterreichische Akademie der Wissenschaften, Technikerstr. 21A, 6020 Innsbruck,
	Austria}
	\affiliation{
	Institut f\"ur Theoretische Physik, Universit\"at Innsbruck,
	Technikerstr. 25, 6020 Innsbruck, Austria}

\author{R. Blatt}
\author{C. F. Roos}
\email{christian.roos@uibk.ac.at}
	\affiliation{Institut f\"ur Quantenoptik und Quanteninformation,\\
	\"Osterreichische Akademie der Wissenschaften, Technikerstr. 21A, 6020 Innsbruck,
	Austria}
	\affiliation{
	Institut f\"ur Experimentalphysik, Universit\"at Innsbruck,
	Technikerstr. 25, 6020 Innsbruck, Austria}

\date{\today}

\maketitle

{\bf
The key to explaining a wide range of quantum phenomena is understanding how entanglement propagates around many-body systems. Furthermore, the controlled distribution of entanglement is of fundamental importance for quantum communication and computation. In many situations, quasiparticles are the carriers of information around a quantum system and are expected to distribute entanglement in a fashion determined by the system interactions \cite{Calabrese2006}. Here we report on the observation of magnon quasiparticle dynamics in a one-dimensional many-body quantum system of trapped ions representing an Ising spin model \cite{Porras:2004,Friedenauer:2008}. Using the ability to tune the effective interaction range \cite{Kim:2009,Britton:2012,Islam:2013}, and to prepare and measure the quantum state at the individual particle level, we observe new quasiparticle phenomena. For the first time, we reveal the entanglement distributed by quasiparticles around a many-body system. Second, for long-range interactions we observe the divergence of quasiparticle velocity and breakdown of the light-cone picture \cite{Lieb1972,Cramer2008,Hastings2010,Nachtergaele2010} that is valid for short-range interactions. Our results will allow experimental studies of a wide range of phenomena, such as quantum transport \cite{Bose2007, guzik_2009_env_assisted}, thermalisation \cite{rigol_relaxation_2007}, localisation \cite{Yao2013} and entanglement growth \cite{Schachenmayer2013}, and represent a first step towards a new quantum-optical regime with on-demand quasiparticles with tunable non-linear interactions.
}

Quasiparticles, such as magnons, phonons, and anyons, are elementary excitations in the collective behaviour of an underlying many-body quantum system. 
While precise control is already possible in the laboratory for systems of individual atoms, ions, or photons, it remains a challenge to extend this to quasiparticles. In systems with nearest-neighbour interactions, quasiparticles are expected to distribute entanglement within light-like-cones defined by a strict quantum information speed limit, enforced not by relativity but by the finite interaction range itself \cite{Lieb1972, Bravyi2006,Eisert2006}.  
These results, known as Lieb-Robinson bounds, have allowed various important theorems to be proven about systems with nearest-neighbour interactions, including restrictions on ground-state correlations \cite{Hastings2006,Nachtergaele2006} and the time to create states for topological quantum computation \cite{Bravyi2006}. Recently, wavefronts of correlations have been observed in bosonic atoms in optical lattices with nearest-neighbour interactions \cite{Cheneau2012,Fukuhara:2013}, and an outstanding challenge is to observe the entanglement dynamics.

\begin{figure}[b]
\begin{center}
\includegraphics[width=0.5\textwidth]{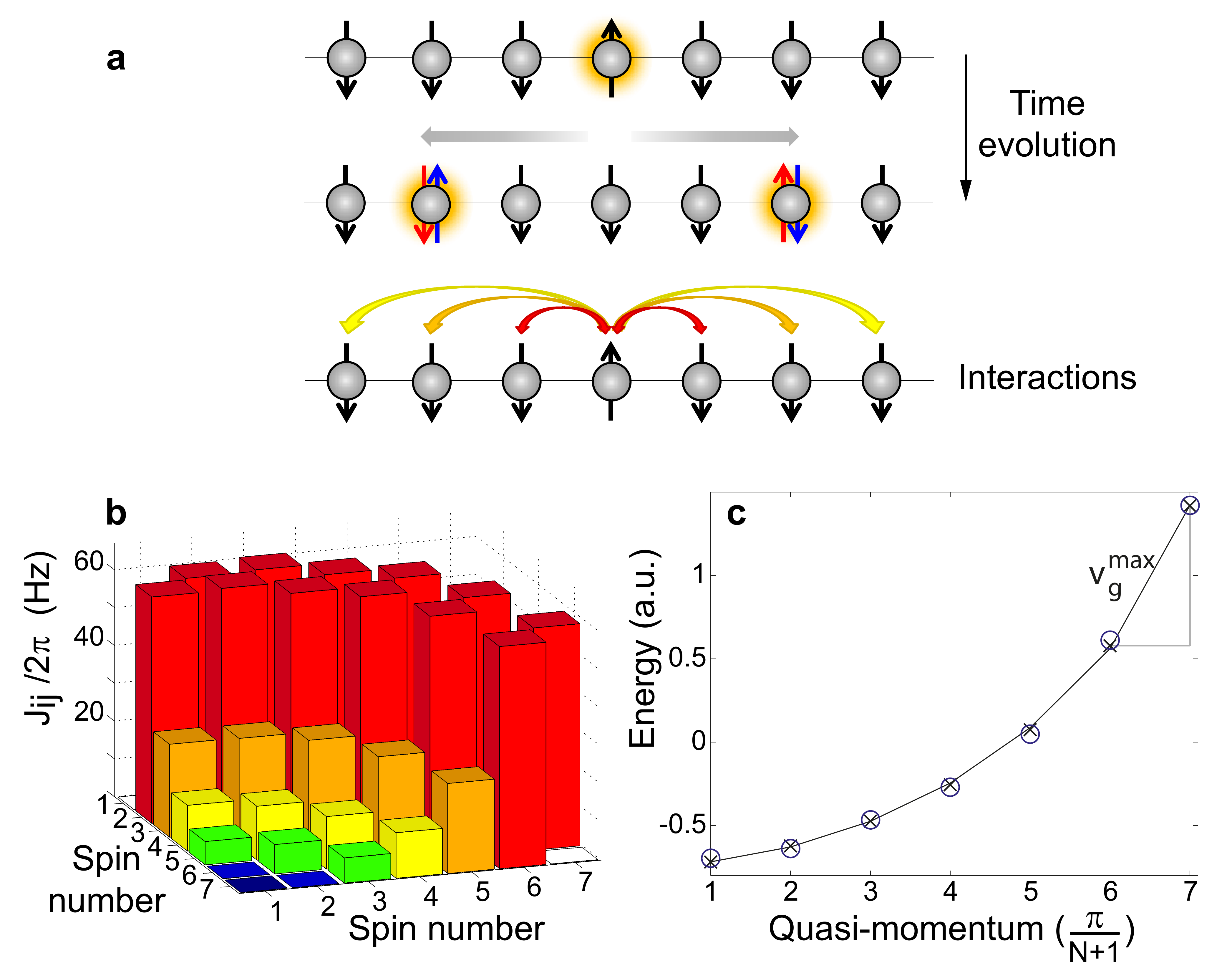}
\vspace{-4mm}
\caption{\label{fig:1b} \textbf{Quantum dynamics in a one-dimensional spin chain following a local quench.} 
\textbf{a}. A steady state is locally perturbed by flipping one of the spins. 
Quasiparticle wave packets propagate left and right from the quench site and entangle spin pairs across the system. The underlying spin--spin interaction defines possible direct hopping paths (examples shown as arrows) and the quasiparticle dispersion relation. 
\textbf{b}. Example long-range spin--spin interaction matrix $J_{ij}$, directly measured in our system of $N=7$ spins (see methods), with colors matched to the interactions pictured in panel a. 
\textbf{c}. Quasiparticle dispersion relation (shifted by energy $\hbar B$), derived from b. (circles) and simulated using experimental parameters (crosses). The line is the fitted dispersion relation for power-law interactions, with best-fit exponent $\alpha=1.36$. The maximum group velocity $v_g^{\rm max}$ is inferred from the curve's steepest slope. Error bars ($1\sigma$) are smaller than symbols used.
}
\end{center}
\vspace{-5mm}
\end{figure}

Extending these results to systems with long-range interactions is of great interest: the interactions in many natural systems fall into this class, exhibiting a power-law dependence ($1/r^{\alpha}$), such as van-der-Waals ($\alpha{=}6$), dipole--dipole ($\alpha{=}3$), or Coulomb interactions ($\alpha{=}1$). In each case, a new set of quasiparticles are predicted with unique properties. If the  interactions fall off sufficiently fast, one can still formulate generalized Lieb--Robinson bounds \cite{Cramer2008,Hastings2010,Nachtergaele2010}. However, the notion of a speed of information propagation becomes invalid. For even longer-range interactions, these bounds break down entirely, which has recently spurred a flurry of theoretical interest \cite{Hauke2013,Schachenmayer2013,Juenemann2013,Eisert2013}. 

We study quasiparticle dynamics in a system of trapped ions which is well described as a one-dimensional chain of interacting spins, with Hamiltonian \cite{Porras:2004,Friedenauer:2008}
\[
H_{\rm Ising} = \hbar\sum_{i<j} J_{ij}\sigma_{i}^x\sigma^x_{j}+\hbar B\sum_i\sigma^z_{i}\,,
\]
where $\sigma^\beta_{i}$ ($\beta=x, y, z$) are the spin-$1/2$ Pauli operators for the $i$'th spin and $B$ is the effective transverse magnetic field strength. The coupling matrix $J_{ij}\sim 1/|i-j|^\alpha$ has approximately a power-law dependence with distance $|i-j|$, with an exponent tunable \cite{Kim:2009,Britton:2012,Islam:2013} between infinite range ($\alpha=0$) and, asymptotically, short range ($\alpha=3$).
We operate in the regime $B\gg\mbox{max}(|J_{ij}|)$, where the number of spin excitations is conserved ($\uparrow_z$) during the dynamics, and $H_{\rm Ising}$ reduces to the $XY$ model of hopping hardcore bosons $H_{XY}=\sum_{i<j}J_{ij}(\sigma_i^+\sigma_j^-+\sigma_i^-\sigma_j^+)$. 

Quasiparticle dynamics can be probed following global or local perturbations (quenches) of a many-body quantum system. Local quenches allow the properties of individual quasiparticles to be studied, i.e., the system's low-lying excitations. In this regime, the system is well approximated by non-interacting quasiparticle modes, each of which corresponds to a single delocalised spin excitation (see methods). We perform a local quench by flipping one of the spins (Fig.~1a), thus coherently populating a broad range of quasiparticle modes. Each of the modes can be pictured as an equal superposition of spin waves (magnons) with positive and negative momentum $\pm k$. Therefore, in the subsequent time evolution, superpositions of left and right travelling spin waves, ejected from the quench site, distribute entanglement amongst the spins. The precise dynamics are determined by the magnon dispersion relation, which in turn can be tuned by the spin--spin interaction range \cite{Hauke2013}. 

In our experiment, spins are encoded in a linear string of 7--15 $^{40}$Ca$^+$ ions. Each ion encodes a spin-1/2 particle in the electronic states $|S_{1/2}, m=1/2\rangle\equiv\spindown$ and $|D_{5/2},m=+5/2\rangle\equiv\spinup$. Spins are manipulated with a narrow-linewidth laser at 729~nm \cite{Schindler:2013}. For further experiment details, see methods.

As a first step, we check the correct implementation of the desired interaction range and the corresponding quasiparticle dispersion relation by directly measuring the exact spatial distribution of the spin--spin interactions (Fig.~1b). The measurement closely matches theoretical predictions. Fitting a dispersion relation from power-law interactions allows us to derive the exponent $\alpha$ (Fig. 1c).
  
Next, we demonstrate the ability to perform arbitrary local quenches by starting with the steady state of all spins down and flipping one or more spins. The spread of information from the quench site(s) can be seen in spatially and temporally resolved single-spin observables like the magnetisation $\langle \sigma_i^z(t)\rangle$.   
 At early times, localised spin-wave packets radiating away from single spin-excitations are clearly visible. At later times, reflections result in complex interference patterns, with properties determined by both $\alpha$ and the spin chain length (Fig.~2a-b). 
Questions, such as whether the initially perfectly localised excitation refocuses, are non-trivial even in the simpler case of nearest-neighbour interactions \cite{Bose2007}. Flipping several spins, e.g. at both ends of the chain, creates counter-propagating wavefronts, opening the prospect of studying quasiparticle collisions and their dependence on the underlying interaction range (Fig.~2c). Extended Data Fig.~1 shows close agreement with theory. Preparing all $N$ spins initially in $\spindown+\spinup$ realizes a global quench. In this case, the many-body state is in a superposition containing 0 through to $N$ excitations, for which interactions between single-excitation quasiparticle modes can no longer be neglected.  Nevertheless, the distribution of information in the resulting dynamics can be investigated by measuring two-point correlation functions \cite{Cheneau2012,Monroe:2013}. Fig.~2d and the Extended Data Fig.~2 show the spread of correlations across the system.

\begin{figure}[t]
\begin{center}
\vspace{0mm}
\includegraphics[width=0.5\textwidth]{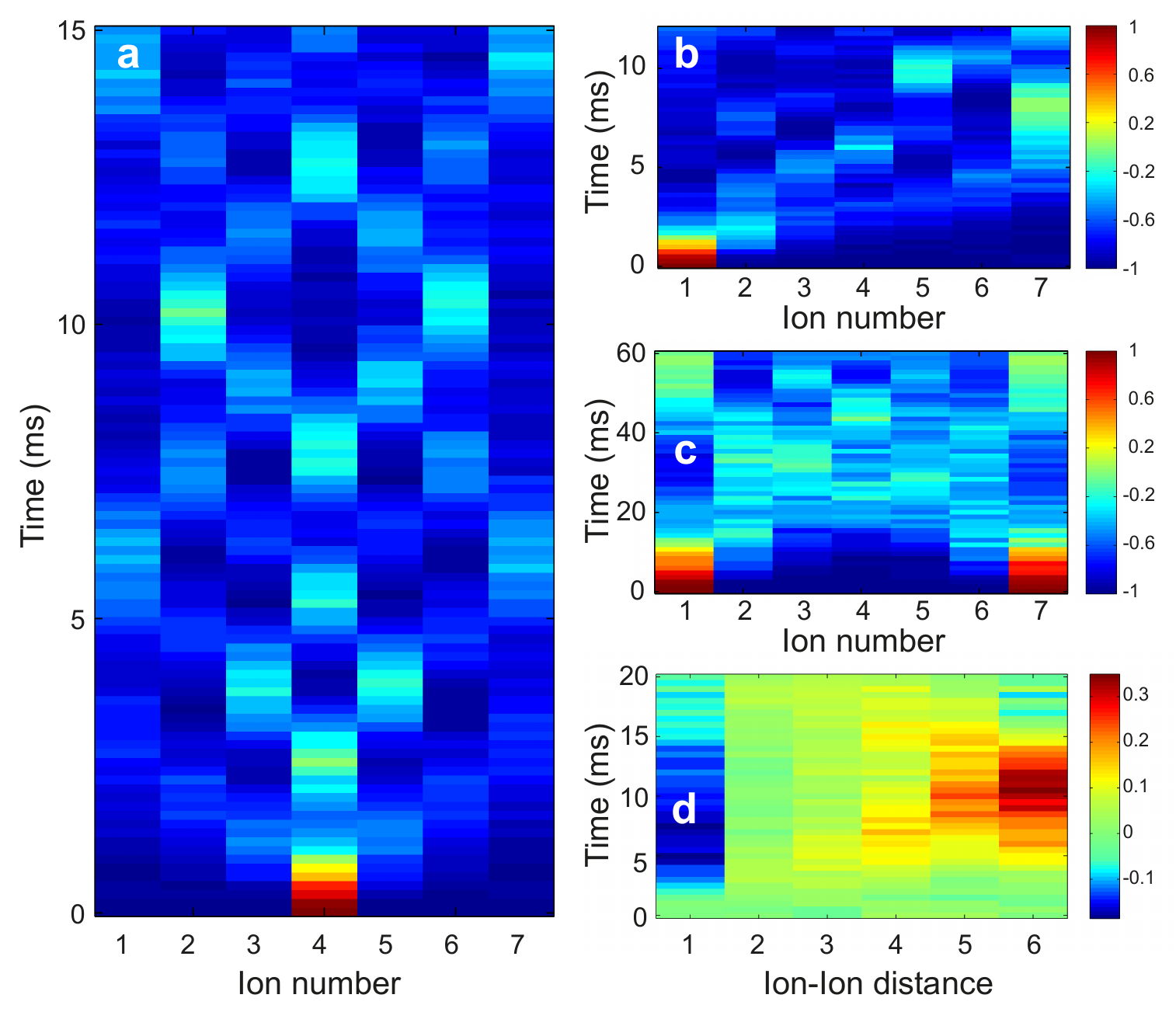}
\vspace{-6mm}
\caption{\label{fig:2} \textbf{Measured quantum dynamics in a systems of seven interacting spins following local and global quenches.} 
\textbf{a-c}. Time evolution of the spatially-resolved magnetisation $\langle \sigma^z_i(t)\rangle$ following a local quench at \textbf{a} the centre spin, for $\alpha~{\approx}~1.36$, \textbf{b} the leftmost spin, for $\alpha~{\approx}~1.36$, and \textbf{c} both ends of the chain, for $\alpha~{\approx}~1.75$.  \textbf{d}. Time evolution of the averaged two-spin correlation function $\overline{C}_n=\frac{1}{N-n}\sum_i^{N-n} C_{i,i+n}$ where $C_{ij}=\langle \sigma^z_i\sigma^z_{j}\rangle -\langle \sigma^z_i\rangle\langle \sigma^z_{j}\rangle$ following a global quench, for $\alpha~{\approx}~1.75$.
}
\end{center}
\end{figure}

We reveal the distribution of quantum correlations after a local quench by tomographically measuring the evolution of the full quantum state of pairs of spins (see Extended Data Fig. 3).
Fig.~3a exemplifies the results for an interaction range $\alpha~{\approx}~1.75$, for which a clear wavefront is apparent. The results show that magnon wave packets emerging from either side of the initial excitation distribute entanglement across the spin-chain (Fig.~3b-c); the wavefront first entangles spins neighbouring the quench site, then the next-nearest neighbours, and so on until the boundaries are reached. 

\begin{figure}
\begin{center}
\includegraphics[width=0.5\textwidth]{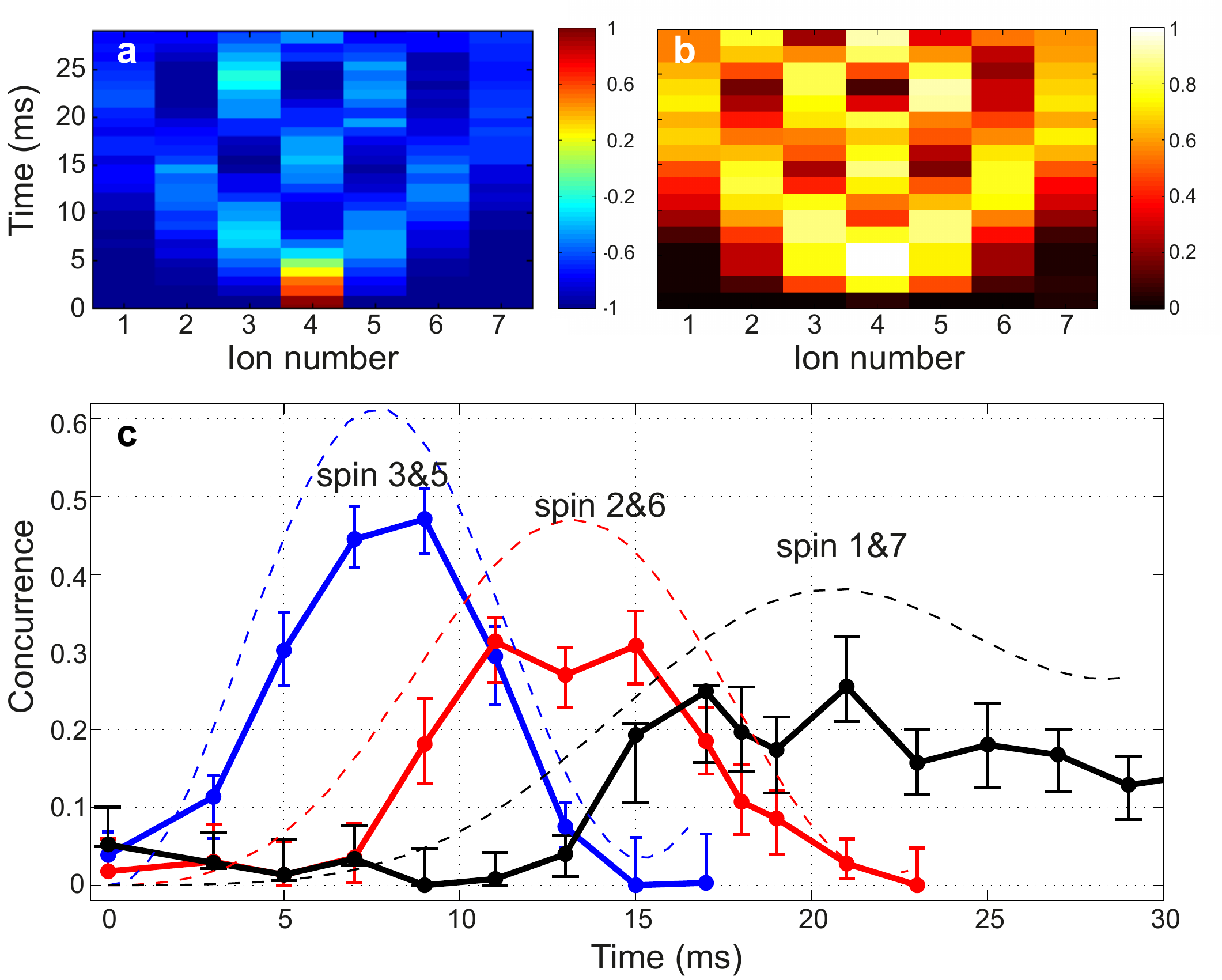}
\caption{\label{label} \textbf{Entanglement distribution following a local quench.} Dynamics for $\alpha~\approx~1.75$. \textbf{a.} Measured single-spin magnetisation \textbf{b.} Single-spin Von Neumann entropy $\mbox{Tr}(\rho\log(\rho))/\mbox{log}(2)$ derived from measured density matrices.  High-entropy states are due to correlations with other spins.  \textbf{c.} Dynamics of entanglement between pairs of spins distributed symmetrically around the central spin, revealing the propagation of entangled quasiparticles from the centre to the boundaries of the system. Blue: spins 3 \& 5. Red: spins 2 \& 6. Black: spins 1 \& 7. Dashed lines show theoretical predictions. $1\sigma$ errors bars are calculated via Monte Carlo simulation of quantum projection noise \cite{Roos:2004}. 
}
\end{center}
\end{figure}

\begin{figure*}[th]
\begin{center}
\includegraphics[width=1\textwidth]{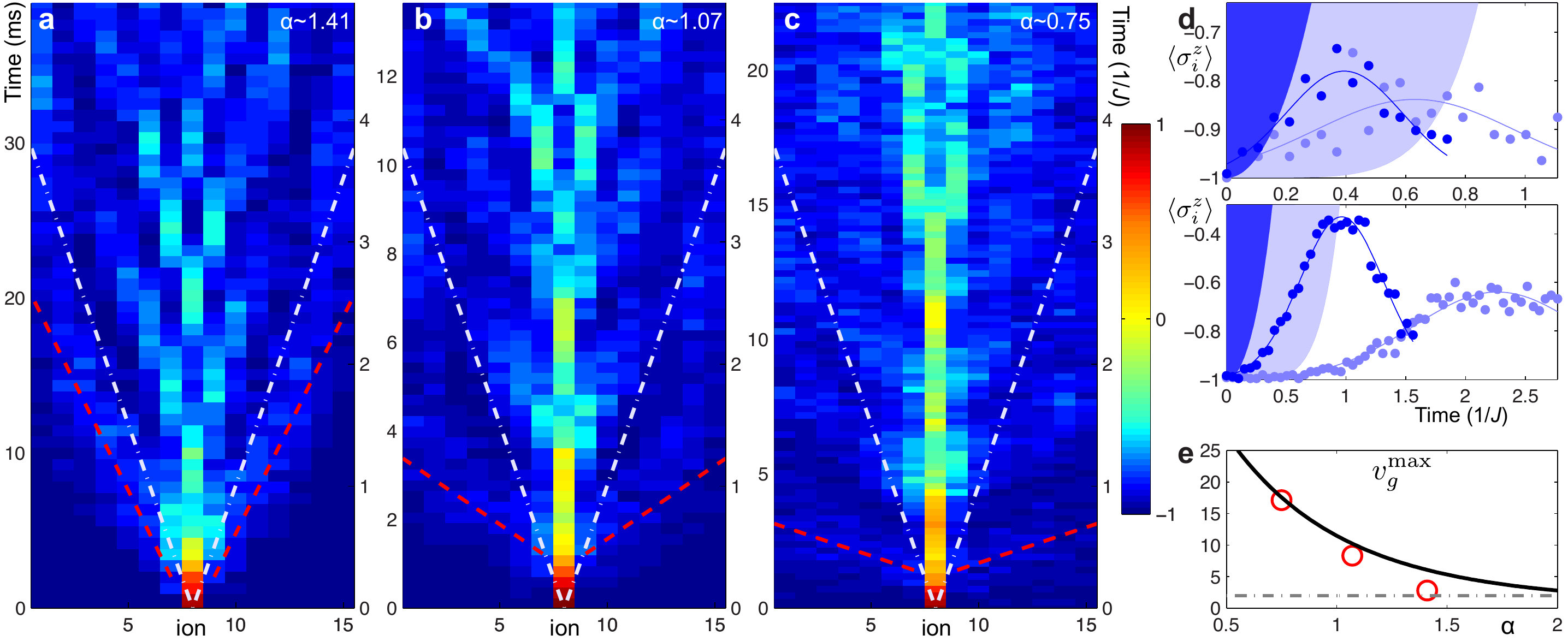}
\caption{\label{label} \textbf{Break-down of the light-cone picture as the underlying interaction range is increased}. \textbf{a-c.} 
In a chain of 15 spins, the central spin is flipped and the subsequent evolution of the magnetisation $\langle \sigma_i^z(t)\rangle$ is measured. 
From a to c, the interaction ranges can be fitted by power laws with $\alpha\sim 1.41,1.07,0.75$. 
Red lines: fits to the experimentally observed magnon arrival times (see methods; examples shown in panel d). 
With increasing interaction range, the signal clearly propagates faster than what is allowed by the nearest-neighbour light-cone (white lines, derived from the maximal group velocity of the nearest-neighbour model). 
\textbf{d.} Magnetisation evolution of spins (ions) 6 and 13, from panel a (bottom) and c (top). 
Solid lines: Gaussian fits to measured magnon arrival. 
Top: In the longest-range case, the signal arrives clearly before what is permitted by the nearest-neighbour Lieb--Robinson bound (shaded region). 
Bottom: In the shortest-range case measured, the slight violation of the nearest-neighbour Lieb--Robinson bound at short times is almost not distinguishable. 
\textbf{e.} Maximum group velocity. 
With increasing $\alpha$, the measured arrival velocities (red circles) of the magnon excitation approach the group velocity of the nearest-neighbour light-cone (grey dash-dotted line). 
For smaller $\alpha$, the measured values are consistent with the divergent behaviour predicted when considering the full power-law interactions (black line).
}
\end{center}
\end{figure*}

Finally we consider how interaction range affects the spreading of information. 
For systems with nearest-neighbour interactions, Lieb and Robinson showed that information cannot propagate away from events, such as local quenches, arbitrarily fast \cite{Lieb1972}:  
observables (e.g. magnetisation) are bounded to an effective light-cone, defined by a line in the space-time diagram beyond which the signal falls off exponentially. The velocity of this light-cone is determined by the nearest-neighbour interaction strength. 

We investigate the validity of the light-cone picture for our system as we vary the interaction range $\alpha$. For this, we carry out local quenches in a chain of 15 spins for three values of $\alpha$, roughly equally spaced around $\alpha=1$. 
In each case, we calculate the effective light-cone that the dynamics should adhere to if only the system's nearest-neighbour interactions contribute. 
In the shortest-range case (Fig.~4a), the arrival times of the first maxima in magnetisation are seen to form a wavefront that is close to the nearest-neighbour light-cone. Cut-throughs in time (Fig.~4d) show a rapid decrease of the measured signal outside this wavefront that is largely captured by the Lieb-Robinson bound (see methods).

As the interaction range is increased (Fig.~4b-c), two things happen. First, the observed wavefronts appear earlier and earlier, reflecting the ejection of faster and faster quasiparticles from the quench site. Second, cut-throughs in time show a gradual fall-off: almost instantly one sees a significant increase in the magnetisation even at large distances (Fig.~4d, top). In this case, information speed is limited not by the interaction range, but by the physical mechanism generating the effective interactions, which in our case is the propagation of acoustic waves across the ion chain \cite{Juenemann2013}. These results clearly show that we can tune our system into a regime in which the physics of short-range interactions do not apply, as predicted by generalised Lieb-Robinson bounds \cite{Hastings2010,Nachtergaele2010}.

We can extract the maximum quasiparticle group velocity  $v_g^{\rm max}$ from the data (see methods and Extended Data Fig.~4). For our shortest range case, the observed  $v_g^{\rm max}$ fits well with the nearest-neighbour case (Fig.~4d). As the interaction range is increased, our results are consistent with a divergence of $v_g^{\rm max}$ as predicted in \cite{Hauke2013}. Note that, despite the faster moving components in the longer-range data (Fig.~4c), the initial perturbation remains more localised. This is consistent with the predicted flattening of the dispersion relation around the divergence. For a comparison of data with theory, see Extended Data Fig.~4.

Differences between the observed and ideal quantum dynamics following local quenches largely correspond to imperfect conservation of excitation number. This could be caused by electric field noise leading to heating 
of the ion's motional state or by unwanted spin-motion entanglement. For global quench dynamics, laser-frequency and magnetic-field fluctuations give rise to dephasing. We anticipate that our experiments can be directly scaled up to tens of spins, and towards regimes where existing numerical and analytical approaches are limited, without changing our approach. 

We have presented a new platform for investigating quantum phenomena---a many-body quantum system in which the states and properties of its quasiparticle excitations can be precisely initialised, controlled, and measured. This immediately opens many new paths for experimental investigations, the subjects of which can be broadly split into the following: 
(1) Quantum transport phenomena, concerning how quantum states and entanglement \cite{Bose2007}, 
or excitations \cite{guzik_2009_env_assisted, plenio_2008_dephasing_transport}, propagate across quantum many-body systems. (2) How quantum systems reach equilibrium, including the question of when thermalisation  \cite{rigol_relaxation_2007, Gong2013} and localisation occur \cite{Yao2013}. (3) Entanglement growth and simulation complexity \cite{Schachenmayer2013} (the interaction range parameter $\alpha$ is known to play a critical role in the growth rate of entanglement and the possibility to simulate the dynamics with conventional computers); 
(4) for quasiparticle behaviour near phase transitions \cite{Sachdev2000}.
For many of these research lines it would be useful, and feasible, to add localised spin excitation absorbers or reflective boundaries, and static or stochastically fluctuating disorder to our system. 

During the final stage of this work, we became aware of complementary recent work using global quenches of trapped-ion spin chains \cite{Monroe:2013}.\\

\noindent {\bf Acknowledgments.} 
We acknowledge inspiring discussions with Luca Tagliacozzo, Markus Heyl and Alexey Gorshkov. 
This work was supported by the Austrian Science Fund (FWF) under the grant number P25354-N20, and by the European Commission via the integrated project SIQS and by the Institut f\"ur Quanteninformation GmbH. We also gratefully acknowledge support from the European Research Council through the CRYTERION Project (\# 227959).\\

\clearpage
\section{Methods Section}
\maketitle
{\bf Magnon eigenmodes.}
The dynamics in our system are well described by an $XY$ Hamiltonian of $N$ coupled spins, 
\begin{equation}
	\label{eq:HXY}
	H_{XY}=\sum_{i<j}J_{ij}(\sigma_i^+\sigma_j^-+\sigma_i^-\sigma_j^+)\,, 
\end{equation}
where $\sigma_i^{\pm}=(\sigma_i^x\pm i \sigma_i^y)/2$, with $\sigma_i^{x,y}$ the usual Pauli matrices. 
We can use the Holstein--Primakoff transformation \cite{Holstein1940} to exactly map the spins to hard-core-interacting bosons, $\sigma_i^+\to a_i^\dagger$, $\sigma_i^-\to a_i$, where $a_i$ ($a_i^\dagger$) annihilates (creates) a boson on site $i$. 
The resulting model, $H_{\rm bos.}=\sum_{i<j}J_{ij}(a_i^\dagger a_j+h.c.)$, conserves the total number of particles (corresponding to conserved magnetisation in the original spin model, allowing us to treat the transverse-field term $B \sum_i \sigma_i^z$ as a constant that we can neglect). 
In the single-particle subspace, diagonalising $H_{\rm bos.}$ is equivalent to diagonalising the $N\times N$ matrix with entries $J_{ij}$. 
The result can be written in the form
\begin{equation}
	H_{\rm bos.}=\sum_{k}\hbar \omega_k a_k^\dagger a_k\,, 
\end{equation}
where $a_k^\dagger=\sum_i c_{i,k} a_i^\dagger$ creates an excitation in eigenmode $k$. The mode functions $c_{i,k}$ are the normalised eigenvectors of $J_{ij}$. 
The eigenmode spectrum $\omega_k$, i.e., the dispersion relation, depends on the boundary conditions and interaction range. Once $J_{ij}$ is known, it can be determined unambiguously.
A single-particle wave-packet constructed from these eigenmodes is what we call a magnon quasiparticle. 

A local perturbation of the fully-polarized state, i.e. a local quench, can be understood as the creation of a single magnon. 
After a spin flip at site $\ell$, e.g. the system state evolves according to 
\begin{equation}
	\label{eq:localQuenchEvolution}
	\ket{\psi(t)} = {\rm e}^{-i\, H_{\rm bos.} t / \hbar} a_\ell^\dagger \ket{0} = \sum_k c_{\ell,k}^*\,\, {\rm e}^{-i\, \omega_k\, t} a_k^\dagger \ket{0}\,,
\end{equation}
where $\ket{0}$ is the vacuum. 
When the number of excitations exceeds one, the picture of non-interacting magnons is only an approximation, and one has to account for the presence of hard-core interactions.

{\bf Calculation of group velocity.}
From Eq.~\eqref{eq:localQuenchEvolution}, it becomes clear that the time evolution after a creation of a single excitation is determined by the magnon dispersion relation, and consequently by the associated group velocities.  
For translationally invariant systems, the modes with energies $\omega_k$ are plane waves with well-defined wave-vector $k$. 
In this case, the magnon group velocities are given by the well-known relation $v_g=\partial \omega/\partial k$. 
In contrast, for a finite system with open boundary conditions and nearest-neighbour interactions, the mode functions are standing waves of the form  $c_i^{(k)}\sim\sin(k\cdot i)$, with $k=n\pi/(N+1)$, $n=1\dots N$. 
In the presence of finite-range interactions, the modes get distorted, but  the number of nodes remains well defined. 
We can then still associate standing waves with the magnon modes and extract a maximal group velocity as $v_g^{\max}\equiv \max_k (\omega_{k+\pi/(N+1)}-\omega_{k}) (N+1)/\pi$.

{\bf Encoding a spin-1/2 in an optical transition of a trapped ion.}
To experimentally realise the spin Hamiltonian $H_{XY}$, we identify the Zeeman states \mbox{$|S_{1/2},m=1/2\rangle$} and \mbox{$|D_{5/2},m^\prime=5/2\rangle$} of $^{40}$Ca$^+$ with the \mbox{$\spindown$} and \mbox{$\spinup$} states of a spin-1/2 particle. The metastable $D_{5/2}$ state has a lifetime of 1.16(2)~s and is connected to the $S_{1/2}$ ground state by an electric quadrupole transition at a wavelength of $\lambda=729$~nm. The degeneracy of the ion's Zeeman states is lifted by a weak magnetic field of $\sim$~4~Gauss which allows us to initialise the \mbox{$\vert S_{1/2},m=1/2\rangle$} state using optical pumping techniques with a probability of about 99.9\%.
Choosing the $|D_{5/2},m^\prime=5/2\rangle$ state for encoding $\spinup$ has the advantage that spontaneous decay of the metastable state does not give rise to population loss from the computational state space. The static electric fields of the linear trap used to confine the ions axially induce electric quadrupole shifts which shift the energy of the \mbox{$\spinup$} state. Hence, the transition frequency $\omega_0$ between the spin states is slightly inhomogeneous across the ion string. However, for our experimental parameters, these inhomogeneities are below $20$~Hz and thus considerably smaller than the spin-spin coupling strength.

{\bf Realisation of variable-range spin-spin couplings in $^{40}$Ca$^+$.}
Variable-range spin-spin interaction of Ising type are realized by globally addressing the ions with a laser beam whose direction is orthogonal to the ion string axis. 
The laser off-resonantly couples the ions' electronic states representing $\spindown$ and $\spinup$ to the ion strings' collective modes of motion in the directions perpendicular to the string. The laser carries two frequencies $\omega_\pm=\omega_0\pm\Delta$ which induces a M{\o}lmer-S{\o}rensen type interaction \cite{Sorensen:1999} by coupling to all first-order sidebands of the transverse collective modes of motion ($\hbar\omega_0$ is the energy difference between $\spindown$ and $\spinup$).  
In the limit of weak coupling, the induced effective interaction between the spins is described by the Hamiltonian \cite{Porras:2004,Kim:2009}
\begin{equation}
H = \hbar\sum_{i<j} J_{ij}\sigma^x_i\sigma^x_j
\label{eq:methods:H}
\end{equation}
with spin-spin coupling constants
\begin{equation}
J_{ij}=\Omega_i\Omega_j\frac{\hbar k^2}{2m}\sum_n\frac{b_{i,n}b_{j,n}}{\Delta^2-\nu_n^2}.
\label{eq:methods:Jij}
\end{equation}
Here, $\Omega_i$ denotes the Rabi frequency of each component of the bichromatic beam on ion $i=1\dots N$, $k=2\pi/\lambda$ and $m$ the ion mass. The summation runs over all $2N$ transverse modes, where $\nu_m$ is the mode's oscillation frequency and $b_{i,n}$ is the Lamb-Dicke factor which is proportional to the displacement of the $i^{th}$ ion in the $n^{th}$ collective mode.

When the laser detuning $\Delta$ is set to a value higher than the frequency of the highest transverse mode, the coupling becomes anti-ferromagnetic with a range which is described approximately by a power-law dependence $J_{ij}\propto |i-j|^{-\alpha}$. The exponent $\alpha$ can be varied between $0$ and $3$. 
The more similar the denominators in eq.~(\ref{eq:methods:Jij}) become, the shorter the range of the interaction gets. This can be achieved by either increasing the laser detuning or by bunching up the transverse modes in frequency space by trapping the ions in a strongly anisotropic potential. In contrast to experiments engineering spin-spin interactions in trapped ions using Raman transitions connecting hyperfine states \cite{Friedenauer:2008, Kim:2009, Britton:2012}, our experiment uses a single-photon transition. 

An additional transverse field $\hbar B\sum_i\sigma_i^z$ can be added to eq.~(\ref{eq:methods:H}) by shifting both components of the bichromatic beam by an additional amount $\delta=2B$. If $B\gg J_{ij}$, joint spin flips coupling \mbox{$\spindown\spindown\leftrightarrow\spinup\spinup$} are suppressed. All local quench experiments presented in the paper were carried out in this regime where the number of excited spins is a conserved quantity. As dephasing due to magnetic field and laser noise is suppressed in subspaces with fixed numbers of excitations, the spin-spin dynamics can be followed over time scales of tens of milliseconds.

In all experiments presented here, with the exception of the data shown in Fig.~4a, we trap ions in a harmonic potential with an axial frequency of $0.219$~MHz and transverse frequencies of $2.655$ and $2.628$~MHz. The degeneracy of the transverse frequencies is slightly lifted to achieve efficient Doppler cooling. The detuning $\Delta$ of the laser from the highest transverse mode is in the range of $15$ to $120$~kHz. To achieve $\alpha=1.41$ in a 15-ion string (Fig.~4a), we lowered the axial confinement to 150~kHz.

To reduce off-resonant excitation of the vibrational modes, frequency resolved sideband cooling of all radial vibrational modes to the ground state is employed at the beginning of each experiment. Rabi frequencies of about $\Omega\approx(2\pi)\,125$~kHz are achieved by focussing about $20$~mW of light to an elliptical Gaussian beam focus with beam waists $w_{||}=380\,\mu$m and $w_\perp=33\,\mu$m. In the seven-ion experiments, the intensity on the outer ions is about 8\% lower than on the central ion. In the fifteen ion experiment, this number increases to about 20\%.

{\bf Compensation of ac-Stark shifts.}
Excitation of the \mbox{$S_{1/2}\leftrightarrow D_{5/2}$} quadrupole transition by the laser inducing the spin-spin interactions gives rise to ac-Stark shifts of the coupled levels. These shifts are caused by off-resonant excitation of dipole transitions coupling the $S_{1/2}$ and $P_{1/2}$ states to other excited states. For our experimental parameters, the light shifts are on the order of 2-3~kHz. Moreover, they vary from one ion to the other, reflecting the intensity inhomogeneities of the laser beam used to the interactions. To compensate the ac-Stark shifts, a third laser frequency is added to the bichromatic beam causing a light shift of the same strength but opposite sign \cite{Haffner:2003}. In order to keep the power of the compensating light field low, we chose to add a frequency component red-detuned by about 1~MHz from the \mbox{$\spindown\leftrightarrow\spinup$} transition. For this method to work, care has to be taken that there are no polarization or k-vector gradients across the ion string that might introduce intensity-independent coupling strength variations among different ions. The intensity of this frequency component is set to the right value by analyzing at which detuning of the bichromatic beam correlated spin flips are observed.

{\bf Single-ion addressing, state tomography and read-out.}
Addressing of single ions is achieved by a strongly focused beam inducing a light shift on one of the ions in the string. The beam position can be switched within $12~\mu$s from one ion to any other ion using an acousto-optic deflector. Arbitrary single ion rotations can be built up from operations combining single-ion ac-Stark shifts with global interactions resonantly coupling the states \mbox{$\spindown$} and \mbox{$\spinup$} \cite{Schindler:2013}. Combining these arbitrary single-ion rotations with spatially resolved detection of the ions' fluorescence on an EMCCD camera enables us to measure any observable that can be written as a tensor product of Pauli spin operators. 

For the measurement of the concurrence of spin-pairs to the left and to the right of the center spin in a chain of seven spins, we carry out quantum state tomography of the respective two-level systems. All required expectation values of two-spin observables $\sigma_i^{\beta_i}\sigma_j^{\beta_j}$ ($\beta=x, y, z$) can be obtained from measurements in nine different measurement bases where ions $1-3$ are projected onto the same set of basis states and ions $4-7$ onto the states of a different measurement basis. In general, it is possible to measure the quantum state of any subset of spins of our system at any point in the dynamics. Error bars for quantities derived from the reconstructed density matrix are obtained from Monte-Carlo bootstrapping techniques \cite{Roos:2004}.

{\bf Measurement of spin-spin coupling matrix elements.}
For the measurement of the spin-spin coupling matrix $J_{ij}$, the ions are initially prepared in $\spindown$ by optical pumping. Next, all ions except $i$ and $j$ are transferred into an auxiliary Zeeman $D_{5/2}$ state, which does not couple to the bichromatic beam inducing the spin-spin coupling, and the state of one of the ions still remaining in \mbox{$\spindown$} is flipped to \mbox{$\spinup$}. Finally, we switch on the bichromatic beam coupling ${\spindown}_i{\spinup}_j$ to ${\spinup}_i{\spindown}_j$ and measure the frequency of oscillation at which the two ions exchange the shared excitation.

{\bf Estimation of $\alpha$ from dispersion relation.}
The spatial behaviour of the spin-spin couplings $J_{ij}$ does not follow a perfect power law, making it difficult to extract an unambiguous exponent $\alpha$ from a direct fit in real space. 
However, the magnon dispersion relation allows us to estimate an effective value for $\alpha$. 
To do this, we compare the dispersion relation of a system with power-law interactions to the dispersion relation of a realistic system obeying the experimental parameters. 
The power-law exponent $\alpha$ yielding the best fit provides an estimate for the interaction range. 
As seen in Fig.~1c, there is a close agreement between the dispersion relation from power-law interactions, the one simulated using experimental parameters, and the one using the measured coupling strengths.
The corresponding $\alpha$ allow us to classify the system behaviour, since the dispersion relation uniquely determines the dynamics in the single-magnon subspace (see Eq.~\ref{eq:localQuenchEvolution}). 
Indeed, as demonstrated in the Extended Data Fig.~1, simulations using the estimated $\alpha$ reproduce the main features of the measured magnetisation dynamics well.

{\bf Numerical simulations of the spin dynamics.}
For numerical simulations, we use the measured trap frequencies and the intensity distribution of the ions across the string to calculate the spin-spin coupling matrix $J_{ij}$. The coupling matrix and the measured laser-ion detuning are then used in a numerical integration of the equation of motion within the $2^N$-dimensional Hilbert space describing the spin system. For the simulation of the fifteen-ion experiments, we disregarded processes not conserving the spin excitation number in order to carry out the numerical integration within the one-excitation subspace. The only free parameter in the simulations is the overall intensity of the bichromatic laser field which could not be perfectly calibrated (deviations from the measured value were always below 5\%).

{\bf Approximate light cones and Lieb--Robinson bound.} 
If interactions in a quantum many-body system are of short range (e.g. exponentially decreasing with distance), information cannot propagate arbitrarily fast. 
Therefore, when measuring an observable (such as the magnetisation $\sigma_i^z$) at a distance $d>0$ from a local perturbation, the observed expectation value can change only after a certain time.  
A mathematically rigorous formulation of this concept, which we will sketch now, has first been given by Lieb and Robinson \cite{Lieb1972} and later generalised by various authors (see, e.g., Refs.~\cite{Hastings2004,Nachtergaele2006,Cramer2008}). 

We denote the unperturbed initial state by $\ket{\psi_0}$ and by $\ket{\psi(t)}$ the state that has evolved during time $t$ following the perturbation. Although the concept is more general, for consistency with Fig.~4 of the main text, we consider a local perturbation with $\sigma_\ell^x$, $\ell=(N+1)/2$, i.e., $\ket{\psi(t=0)}=\sigma_\ell^x \ket{\psi_0}$. 
The change of any observable $\mathcal {O}$ can then be bounded by 
\begin{equation}
|\bra{\psi(t)}\mathcal{O}\ket{\psi(t)}-\bra{\psi_0}\mathcal{O}\ket{\psi_0}| \leq \left|\left|\left[\mathcal{O}(t),\sigma_\ell^x(0)\right]\right|\right|\,,
\end{equation}
where $\mathcal{O}(t)$ is the observable $\mathcal{O}$ evolved in the Heisenberg picture of the unperturbed Hamiltonian, and where $||\mathcal O||$ is the operator norm of $\mathcal O$, i.e. its largest absolute value of its eigenvalues. 
The commutator on the right hand side quantifies how much it matters in which temporal order the operators $\mathcal{O}$ and $\sigma_\ell^x$ are applied. 

If interactions decrease exponentially with distance, one can bound this commutator by \cite{Hastings2010,Nachtergaele2010} 
\begin{equation}
	\label{eq:LRbound}
	\left|\left|\left[\mathcal{O}(t),\sigma_\ell^x(0)\right]\right|\right| \leq \, \left|\left|\mathcal{O}\right|\right|\, F(d,t)\,,\quad F(d,t)=C\, \ue^{\,\mu\left(v|t|-d\right)}\,, 
\end{equation}
where $C$ and $\mu$ are positive constants that depend on the interactions and lattice structure, and $v$ is the so called Lieb--Robinson velocity. 
In its essence, the function $F(d,t)$ provides an approximate light cone -- information propagating faster than $v$ is exponentially suppressed.  
The appearing constants are not unique  \cite{Hastings2010,Nachtergaele2010}, but as a reasonable choice one may identify the Lieb--Robinson velocity with the maximal group velocity $v_g^{\rm max}$.   
To study the break down of the light-cone picture in systems with long-range interactions, we include in Fig.~4(a-c) of the main text the line $t=d/v_g^{\rm max}$ that would bound the approximate light cone if the system had only nearest-neighbours interactions with strength $\bar J\equiv\frac{1}{N-1}\sum_{i=1}^{N-1} J_{i,i+1}$. 
The initial excitation clearly propagates faster than this nearest-neighbour light cone. 

For the case of nearest-neighbour interactions, one can find a compact formulation of the Lieb--Robinson bound~\eqref{eq:LRbound}, which we use in Fig.~4d to quantify its violation in long-range interacting systems. 
Assume a nearest-neighbor Hamiltonian $H=\sum_{ij}h_{ij}$. Then, following Ref.~\cite{Bravyi2006}, one can write 
$F(d,t)=\sum_{m=d}^\infty \mathcal{N}(m) {(2 g |t|/\hbar)^m}/{m!}\,$, with $g={\max}_{ij} ||h_{ij}||$. Here, $\mathcal{N}(m)$ is the number of paths with length $m$ that connect the quenched site $\ell$ to the observable $\mathcal{O}$ at distance $d$. 
In one dimension, simple counting gives 
$\mathcal{N}(m)=\left(\begin{array}{c} m \\ \frac{m-d}{2} \end{array}\right)$ if either $m$ and $d$ are both even or both odd, and $\mathcal{N}(m)=0$ otherwise. 
Using this property, one can analytically evaluate the sum in $F(d,t)$ which takes the particularly elegant form 
\begin{equation}
	\label{eq:LRbound}
	\left|\left|\left[\mathcal{O}(t),\sigma_\ell^x(0)\right]\right|\right| \leq \, \left|\left|\mathcal{O}\right|\right|\, F(d,t)\,,\quad F(d,t)=2 I_d(4 g |t|/\hbar)\,, 
\end{equation}
where $I_d(x)$ is the modified Bessel function of the first kind. 
In Fig.~4d of the main text, we study the corresponding bound for $\mathcal{O}=\sigma_i^z$ and $g=\max_i J_{i,i+1}$, showing that the signal in our system propagates faster than allowed by the nearest-neighbour Lieb--Robinson bound.\\

\bibliographystyle{naturemag}
\bibliography{bibliography2}

\clearpage
\section{Extended data figures}
\renewcommand{\figurename}{Extended Data Figure}
\setcounter{figure}{0}

\begin{figure}[htp]
\begin{center}
\includegraphics[width=0.5\textwidth]{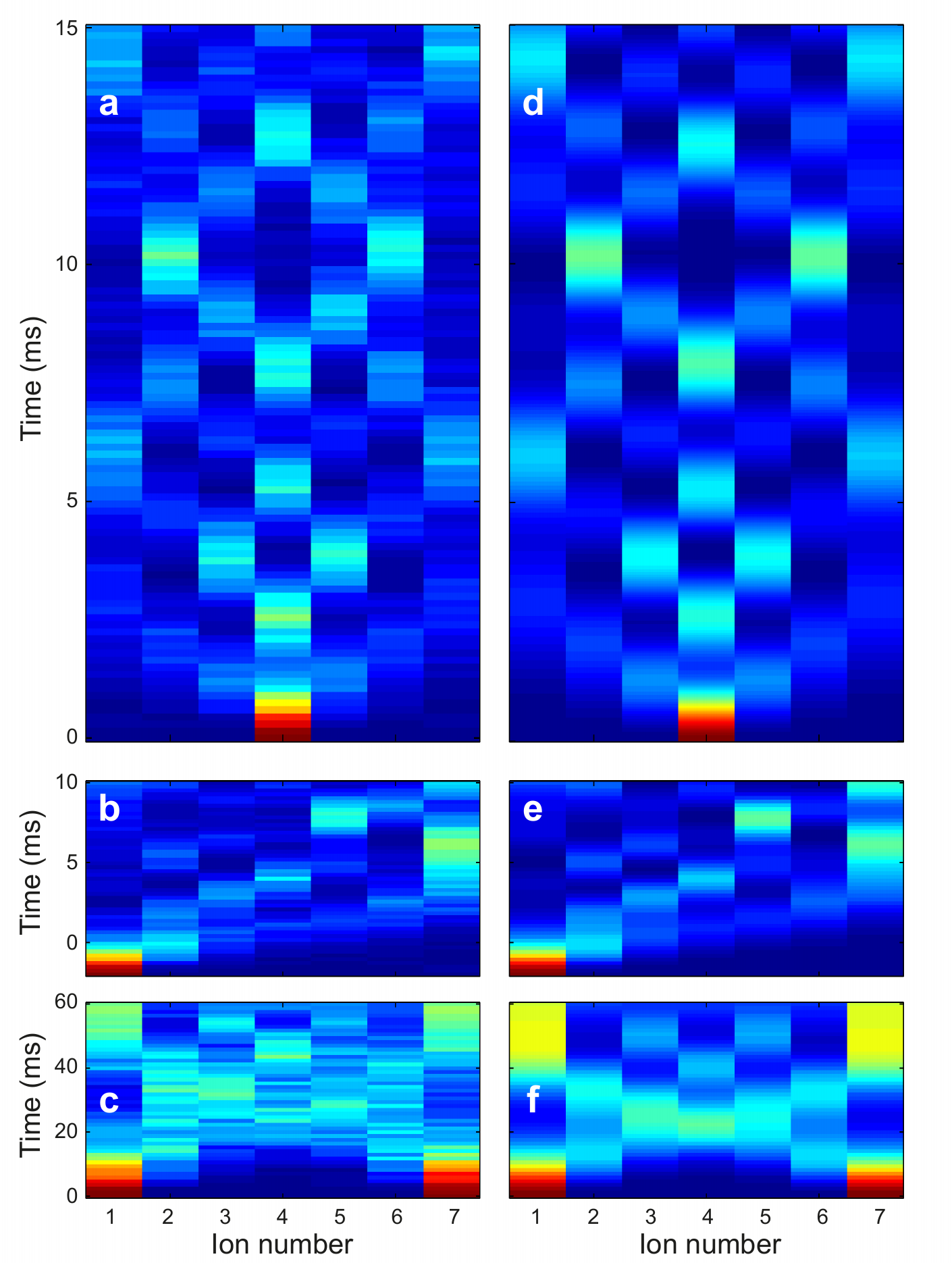}
\vspace{-5mm}
\caption{\label{labelss} \textbf{Quantum dynamics following local quenches in a seven spin system.} Time evolution of the spatially-resolved magnetisation $\langle \sigma^z_i(t)\rangle$ for three different local quenches. In each panel measured data (LHS) is shown next to theoretical calculations for the ideal case (RHS). 
\textbf{a}. quench at the centre spin, $\alpha~{\approx}~1.36$. 
\textbf{b}. quench the leftmost spin, $\alpha~{\approx}~1.36$. 
\textbf{c}. quench at both ends of the chain, $\alpha~{\approx}~1.75$. 
Theoretical calculations employ measured laser-ion coupling strengths and distribution across the ion chain.
}
\end{center}
\end{figure}

\begin{figure}[t]
\begin{center}
\vspace{-5mm}
\includegraphics[width=0.5\textwidth]{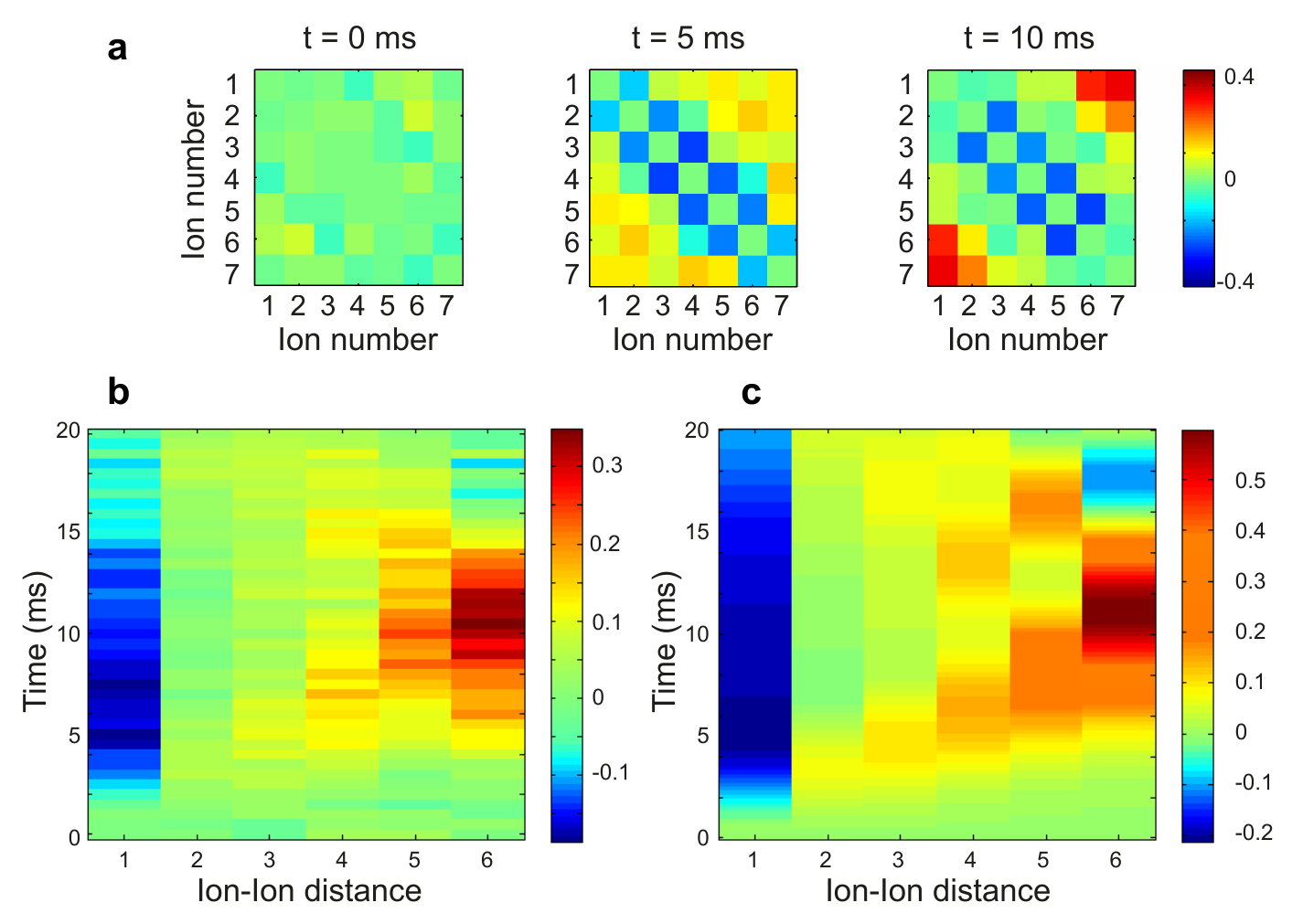}
\vspace{-5mm}
\caption{\label{labelss} \textbf{Quantum dynamics following a global quench in a seven spin system.} 
\textbf{a}. Measured correlation matrices with elements $C_{ij}(t)=\langle \sigma^z_i(t)\sigma^z_j(t)\rangle -\langle \sigma^z_i(t)\rangle\langle \sigma^z_j(t)\rangle$ at t = 0~ms, 5~ms, and 10~ms ($\alpha~{\approx}~1.75$). 
\textbf{b}. Measured average spin-spin correlations $\overline{C}_n(t)=\frac{1}{N-n}\sum_i^{N-n} C_{i,i+n}(t)$ as a function of time and distance $n$ where the average was taken over all spin pairs $(i,j)$ with $|i-j|=n$.
\textbf{c}. Calculated spin-spin correlations  $\overline{C}_n(t)$ as a function of time and distance (notice the different colorbar). 
}
\end{center}
\end{figure}

\begin{figure}[bh]
\begin{center}
\vspace{-3mm}
\includegraphics[width=0.5\textwidth]{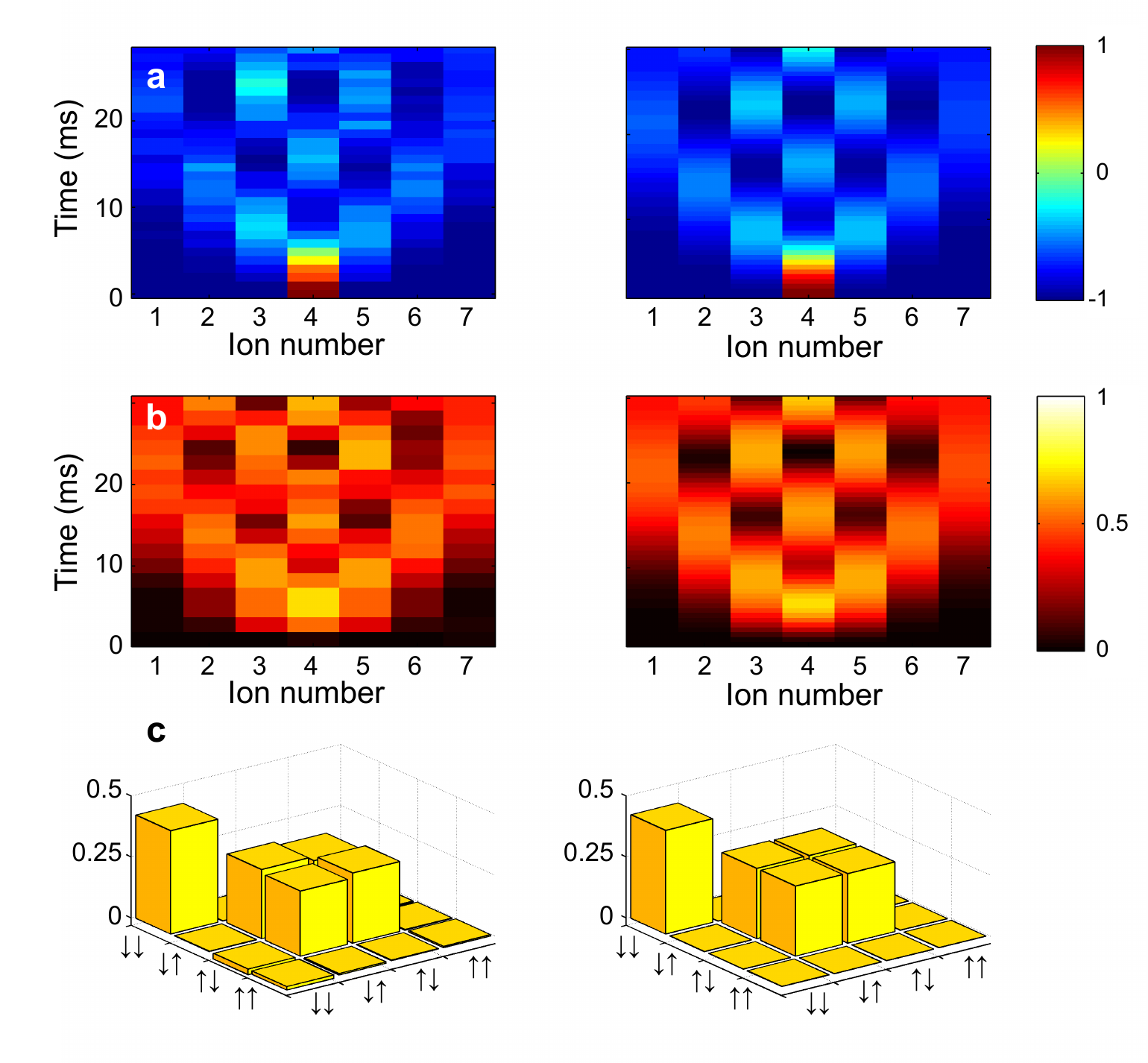}
\vspace{-10mm}
\caption{\label{labelss} \textbf{Entanglement distributed by quasiparticles.} Following a quench of the central spin with short-range interactions ($\alpha~\approx~1.75$) a distinct wavefront emerges. 
Each panel shows data (LHS) and theory (RHS)
\textbf{a.} measured single-spin magnetisation. 
\textbf{b.} Single-spin Von Neumann entropy $\mbox{Tr}(\rho\log(\rho))$ normalized to 1, derived from measured density matrices. 0 would correspond to a fully pure quantum state (black) and 1 a fully mixed state. The increase in entropy of any individual spin during the dynamics reflects the generation of entanglement with other spins. 
\textbf{c.} Real part of the tomographically reconstructed full density matrix of spins 3 and 5 at a time 9~ms after the quench. Imaginary parts are less than 0.03. The fidelity between the full experimentally reconstructed $\rho$ and ideal state $\ket{\psi}$ is $F=0.975\pm{0.005}$, using $F=Tr(\rho\ket{\psi}\bra{\psi})$.
}
\end{center}
\end{figure}

\begin{figure*}[th]
\begin{center}
\includegraphics[width=0.8\textwidth]{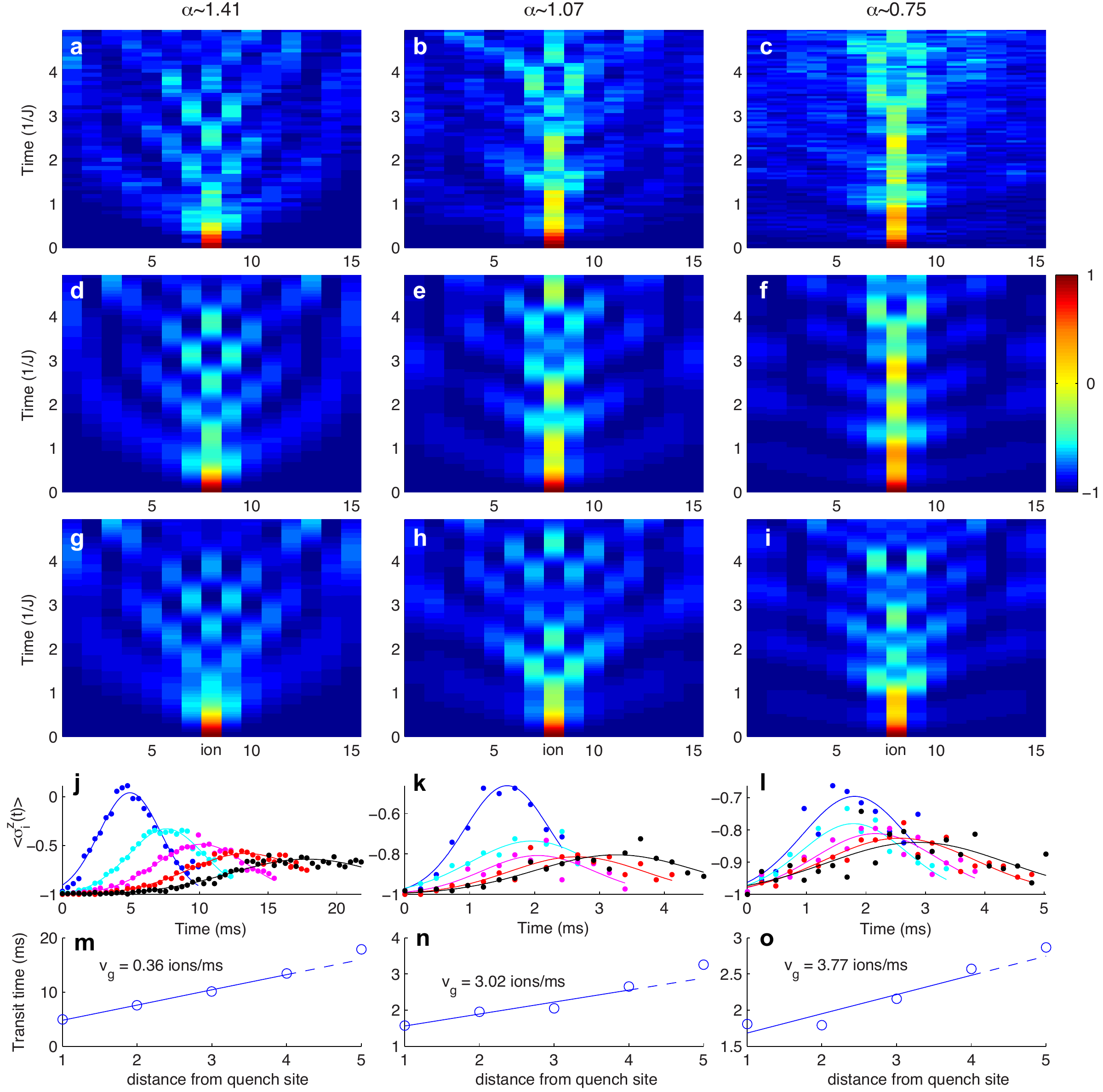}
\caption{\label{labelss} \textbf{Quantum dynamics following a local quench in a fifteen spin system.} 
{\bf a-c} Experimentally measured time evolution of $\langle \sigma^z_i(t)\rangle$ [as Fig.~4(a-c)]. 
{\bf d-f} Theoretical calculations based on measured spin--spin interaction matrix (such as presented in Fig.~1b).
{\bf h-i} Theoretical calculations using $J_{ij}=\bar J/|i-j|^\alpha$, with $\bar J=\frac{1}{N-1}\sum_{i=1}^{N-1}J_{i,i+1}$ and $\alpha$ extracted from a fit to the measured dispersion relation. 
All theory calculations are done in the single-excitation subspace. 
The agreement between theory and experiment is excellent. 
{\bf j-l} Gaussian fits to the measured arrival time of the first quasiparticle maximum (from panels a-c). 
{\bf m-o} Excluding the outermost ion to reduce finite-size effects, the fitted arrival maxima (circles) trace approximately a straight line when plotted against distance. A linear fit (solid line) yields an estimate for the propagation speed of the first quasiparticle maximum. 
}
\end{center}
\end{figure*}

\end{document}